\documentclass[conference]{IEEEtran}

\usepackage{amsmath}
\usepackage{amsfonts}
\usepackage{amssymb}
\usepackage{amsmath}
\usepackage{cite}
\usepackage{subfig}
\usepackage{graphicx}
\usepackage{algorithm,algcompatible}
\algnewcommand\INPUT{\item[\textbf{Input:}]}%
\algnewcommand\OUTPUT{\item[\textbf{Output:}]}%
\ifCLASSINFOpdf
\else
\fi

\begin{document}
\title{Spectrum Sharing by Space-Time Waveform Shaping \thanks{This work was supported in part by the National Science Foundation
under Grants CNS-2027138, CNS-2117822 and EEC-2133516 and the Air
Force Research Laboratory under Grant FA8750-21-1-0500. Distribution A. Approved for public release: Distribution Unlimited: AFRL-2024-6086 on 30 Oct 2024.}}

\author{\IEEEauthorblockN{ 
    Hatef Nouri$^*$, George Sklivanitis$^*$, Dimitris A. Pados$^*$, and Elizabeth Serena Bentley$^{\dagger}$}
\IEEEauthorblockA{$^*${Center for Connected Autonomy and AI, Florida Atlantic University,} Boca Raton, FL 33431 USA\\ 
E-mail: \{hnouri, gsklivanitis, dpados\}@fau.edu\\
$^{\dagger}$Air Force Research Laboratory, Rome, NY 13440 USA}
E-mail: elizabeth.bentley.3@us.af.mil
}
% the acknowledgment of grants, issue a 
\IEEEoverridecommandlockouts

%\author{\IEEEauthorblockN{ 
%    Hatef Nouri$^*$, George Sklivanitis$^*$, Dimitris A. Pados$^*$, and Elizabeth Serena Bentley$^{\dagger}$}
%\IEEEauthorblockA{$^*${Center for Connected Autonomy and AI, Florida Atlantic %University,} Boca Raton, FL 33431 USA\\ 
%$^*$E-mail: \{hnouri, gsklivanitis, dpados\}@fau.edu\\
%$^{\dagger}$Air Force Research Laboratory, Rome, NY 13440, USA}
%$^{\dagger}$E-mail: \{elizabeth.bentley.3\}@us.af.mil
%}

\maketitle

\linespread{0.905}
\begin{abstract}
In this paper, we consider the task of introducing a new wireless data link over a given occupied frequency band using a multi-antenna transmitter and receiver.
We design formally a dynamic multiple-input multiple-output (MIMO) wireless link that can coexist in the fixed congested frequency band by (a) optimally avoiding sensed interference in the joint space-time domain, and (b) protecting existing links by minimizing its own transmitted power in the band. In particular, the transmit beam weight vector and time domain pulse code sequence are jointly optimized to minimize the transmit energy per bit per antenna, while maintaining a pre-defined signal-to-interference-plus-noise ratio (SINR) at the output of the joint space-time maximum SINR receiver filter. Extensive numerical studies are carried out to demonstrate the derived algorithmic solution in 
%light %and dense occupied 
light and heavily congested band scenarios with non-cooperative co-channel links. We show that the proposed autonomously reconfigurable 4x4 MIMO link outperforms a non-adaptive transceiver and other forms of waveform shaping in terms of the pre-detection SINR performance and the capability to protect ongoing non-cooperative links by not occupying the band with redundant transmissions. 

%which optimally shapes the transmitted waveform  reception jointly in time and space to coexist in a utilized frequency band in two manners: (a) optimally avoid sensed existed interference and disturbance by cognitive reshaping of the waveform and (b) optimally protect the neighbor links by minimizing the transmitted power in the band. The cognition is based on the estimated channel coefficients and occupancy autocorrelation matrix. In particular, beam weight vector and the pulse code sequence are adjusted to minimize the transmit energy per bit and per antenna while maintaining a maximum pre-defined signal-to-interference-plus-noise ratio (SINR). We formulate the design of cognitive MIMO transceiver and extensive numerical results are provided to demonstrate derived algorithmic solutions under spread-spectrum/non-spread-spectrum interference, in light and dense interference scenarios. Further, the designed transceiver is implemented on RFSoC platforms. Our transceiver dynamically optimizes beam weight vector and the pulse code sequence to shape the waveform and transmitted energy. In the presence of different levels of co-channel interference, survivability and protectability of our link is demonstrated.
\end{abstract}

%\begin{IEEEkeywords}
%Frequency coexistence, interference avoidance, signal-to-interference-plus-%noise ratio, space-time waveform design, spectrum sharing.
%\end{IEEEkeywords}

\IEEEpeerreviewmaketitle

\section{Introduction}

\IEEEPARstart{F}{requency} coexistence problems emerge as key challenges in the realm of modern wireless communication systems as ever expanding collections of devices compete for access to limited frequency spectral bands resulting in interference, signal degradation, and compromised data integrity \cite{ref1}. The spectrum sharing predicament is especially pronounced in densely occupied areas where numerous wireless technologies, such as machine-to-machine communications \cite{ref2}, Internet-of-Things \cite{ref3}, mm-wave robotics \cite{ref4}, and others, coexist within close physical proximity. Addressing this challenge requires sophisticated spectrum management strategies \cite{akyildiz2008survey}, advanced signal processing techniques \cite{hamalainen2002uwb}, and collaborative efforts to sustain comfortable coexistence of diverse wireless technologies on shared frequency bands \cite{naik2018coexistence}. 

At a high level of abstraction, spectral occupancy and interference is broadly handled by licensing.
%, addressing occupancy and interference issues might be a solution to free up some of the licensed spectrum conditioned on availability of cognitive non-occupancy protocols. For seamless frequency coexistence, it is crucial to develop a
Today we understand that adaptive radio systems capable of performing real-time waveform shaping (in msec time scale or better) can lead to improved utilization of both licensed and unlicensed spectral bands \cite{sklivanitis2015all,tountas2018all,ding2013all,sklivanitis2017sparse,sklivanitis2018airborne}. Indeed,
in recent years there has been a growing interest in the research literature on spectrum sharing strategies to navigate the complexities of the contemporary wireless environment \cite{lee2008optimal,kang2010optimal,rose2002wireless}. Wireless systems of the near future will have elements which adapt dynamically to changing patterns of interference by adjusting modulation and signal processing methods in much the same way that power control \cite{yates1995framework} is used today. An approach to maintain wireless connectivity in overloaded network setups is described in \cite{tountas2019dynamic} where an optimal algorithm is proposed to adaptively design sparse waveforms with well-placed energy that maximize the signal-to-interference-plus-noise ratio (SINR) at the output of the maximum-SINR linear filter at the receiver side. On the other hand, multiple-input multiple-output (MIMO) technology, which is a standard component in this and next generations of communication systems \cite{zhang2020prospective}, presents unique opportunities to ameliorate the problem of spectral co-existence by introducing degrees of freedom in the space domain. 
In \cite{abdulkadir2016space}, directional transmission/reception and space-time precoding/filtering is offering flexibility in waveform shaping and co-existence scenaria \cite{abdulkadir2016space}. Directional space-time waveform design for MIMO configuration is discussed in \cite{tountas2019directional}, in which code sequence and signal angle-of-arrival (AoA) are jointly optimized to maximize the attainable SINR for proactive interference avoidance. In \cite{liu2012joint}, a precoder is proposed to jointly suppress multiuser interference and other-cell interference. In \cite{zhang2008exploiting}, multi-antennas at the secondary transmitter are exploited to effectively balance between spatial multiplexing for the secondary transmission and interference avoidance at the primary receivers. The trade-off is studied from an information-theoretic perspective by characterizing the secondary user's channel capacity under both its own transmit-power constraint as well as a set of interference-power constraints each imposed at one of the primary receivers. The work in \cite{osman2021novel} proposes an interference avoidance distributed deep learning model for IoT and device-to-any-destination communication.

In this paper, we design a new wireless data link to reside in any occupied fixed frequency band by optimally avoiding sensed interference and protecting existing in-band transmissions.
We address formally the fundamental problem of optimizing a MIMO link to achieve a given desired receiver SINR with minimum transmit power. The proposed solution takes the form of optimal dynamic joint-space-time shaping of the transmit waveform  for any locally sensed occupancy autocorrelation matrix upon joint-space-time signal reception.
%the modelling of a dynamically cognitive MIMO wireless transceiver which optimally shapes the transmitted waveform and reception jointly in time and space to avoid sensed existed interference and disturbance by cognitive reshaping of the waveform and minimize the transmitted power in the band. Beam weight vector, pulse code sequence, and transmit energy per bit and per antenna are optimally adjusted while maintaining a maximum pre-defined pre-detection SINR at the output of the joint space-time receiver filter for any locally sensed space-time occupancy autocorrelation matrix. We formulate the design of cognitive MIMO transceiver and propose a novel closed-loop transmit space-time signal design solution to dynamically optimize the transmitter beam weights, code sequence vector, and transmit energy per bit and per antenna. Extensive numerical results are provided to demonstrate derived algorithmic solutions under spread-spectrum/non-spread-spectrum interference, in light and dense interference scenarios. 

The rest of the paper is organized as follows. In Section II, we present the MIMO signal model. In Section III, we formulate the sensing and optimization problem. In Section IV, we present the proposed optimum waveform design. Some conclusions are drawn in Section V.

\emph{Notation}: ${{\left( \text{ }.\text{ } \right)}^{T}}$, ${{\left( \text{ }.\text{ } \right)}^{*}} $, and ${{\left( \text{ }.\text{ } \right)}^{H}}$ denote transpose, conjugate, and Hermitian operations, respectively; $\overline{f\left( t \right)}$ represents the time average of signal $f\left( t \right)$. Bold upper-case and lower-case letters denote matrices and column vectors, respectively. The ${{\left( i,j \right)}{th}}$ entry of a matrix $\mathbf{A}$ is denoted by ${{a}_{ij}}$ or ${{\left[ \mathbf{A} \right]}_{ij}}$. $E\{ \text{ }.\text{ } \}$ denotes statistical expectation; $\mathbb{R}$ is the set of real numbers and ${{\mathbb{Z}}^{+}}$  the set of non-negative integers; ${{\mathbf{I}}_{N}}$ is the $N\times N$ identity matrix;
%$X\sim\mathcal{N}\left( \mu ,{{\sigma }^{2}} \right)$ denotes Gaussian random variable with mean $\mu $ and variance ${{\sigma }^{2}}$.
$diag\left\{ {{x}_{1}},{{x}_{2}},...,{{x}_{M}} \right\}$ is the diagonal $M\times M$ matrix with elements ${{x}_{i}}$, $i = 1,2,...,M$, on the diagonal; the Kronecker product of two matrices/vectors is denoted by $\otimes $.

\section{MIMO System Model}

We consider a generic MIMO system with $M$ transmit and $N$ receive antennas all in arbitrary formation capable of establishing a wireless communication link with bandwidth $BW$ around center carrier frequency $f_c$.

%over a frequency-flat Rayleigh fading channel. Under the assumption of slow fading, we assume that the channel coefficients remain constant over a block of $B$ symbols, i.e., $B$ is much smaller than the channel coherence time.
%\subsection{Signal Model}

We assume deployment of an arbitrary $Q$-ary digital modulation scheme -for example, quadrature-amplitude-modulation (QAM)- under which the $m$th antenna on the transmitter side emits
\begin{equation}
{{x}_{m}}\left( t \right)=\sum\limits_{k=0}^{K-1}{\sqrt{{{E}_{s}}}b\left[ k \right]s\left( t-k{{T}_{s}} \right){{e}^{j2\pi {{f}_{c}}t}}{{w}_{m}}},
\label{eq1}
\end{equation}
$m=1,2,...,M$, where $w_m \in \mathbb{C}$ is a gain-and-phase adjustable antenna parameter and the information symbols 
$b\left[ k \right]\in \left\{ {{b}_{1}},{{b}_{2}},...,{{b}_{Q}} \right\}$, $k=0,1,...,K-1$, reside on a digitally shaped waveform $s\left( t \right)$ of duration $T_s$,
\begin{equation}
s\left( t \right)=\sum\limits_{l=0}^{L-1}{{{s}_{l}}p\left( t-l{{T}_{c}} \right)}
\label{eq2},
\end{equation}
created by $L \geq 1$ coded repeats, ${{s}_{l}}\in \left\{ \pm 1/\sqrt{L},\pm j/\sqrt{L} \right\}$, $l=0, ...,L-1$, of a square-root raised cosine (SRRC) pulse
 $p\left(\cdot\right)$  with roll-off factor $\alpha $ and duration ${{T}_{c}}$  (i.e., ${{T}_{s}}=L{{T}_{c}}$). 
 In the model of (\ref{eq1}), (\ref{eq2}), 
 the bandwidth of the transmitted signal by each antenna  is $BW=\left( 1+\alpha  \right)/{{T}_{c}}$ and each symbol is transmitted by all antennas $m=1,...,M$ creating a MIMO link data rate of $1/{{T}_{s}}$ symbols/sec (or  ${{\log }_{2}}Q/{{T}_{s}}$ bits/sec.) Assuming that the energy of the individual pulse is unity, i.e., $\int_{0}^{{{T}_{c}}}{{{\left| p\left( t \right) \right|}^{2}}dt}=1$, the total transmit energy per symbol for the MIMO link is
\begin{equation}
{{E}_{T}}={{E}_{s}}\sum\limits_{m=1}^{M}{{{\left| {{w}_{m}} \right|}^{2}}}\label{ET}.
\end{equation}
%where $w_m \in {\mathbb{C}}$ is the gain and magnitude
%adjustable 
%beam weight parameter for antenna $m$, $m=1,...,M$, in (\ref{eq1}).
Enforcing a norm constraint on the beam weight vector $\mathbf{w} \triangleq [ w_1, \, w_2, \, \ldots, w_M]^T \in {{\mathbb{C}}^{M\times 1}}$ of the form ${{\left\| \mathbf{w} \right\|}^{2}}=M$
implies that setting each transmit antenna ${{E}_{s}}={{E}_{T}}/M$  results in total system transmit energy per symbol equal to $E_T$.

At the receiver side, after frequency down-conversion, the signal received by the ${n}$th antenna element is modeled by
\begin{equation}
\begin{split}
{{y}_{n}}\left( t \right)=&\sum\limits_{k=0}^{K-1}
\sum\limits_{m=1}^{M}{{{h}_{m,n}}{{w}_{m}}}
{ \sqrt{{{E}_{s}}}b\left[ k \right]s\left( t-k{{T}_{s}} \right) } \\
&+{{i}_{n}}\left( t \right)+{{\omega }_{n}}\left( t \right), \,\,\, n=1, ..., N,
\end{split}
\end{equation}
where ${{h}_{m,n}} \in \mathbb{C}$ represents the $m$th-to-$n$th-antenna channel coefficient that is assumed to remain constant during the duration of the $K$ transmitted symbols, and ${{i}_{n}}\left( t \right)$, ${{\omega }_{n}}\left( t \right)$ 
capture compound interference and additive white Gaussian noise, respectively, experienced by the ${n}$th receive antenna. After individual pulse-matched filtering over $L$ pulses corresponding to transmitted symbol $k$ at each receive antenna element $n$, the complete symbol decision statistic is comprised of the values
\begin{equation}
\begin{split}
&{{y}_{n,l}}\left[ k \right]=\sum\limits_{m=1}^{M}{ {{h}_{m,n}}{{w}_{m}}}
\sqrt{{{E}_{s}}}b\left[ k \right]{{s}_{l}}+{{i}_{n,l}}\left[ k \right]+{{\omega }_{n,l}}\left[ k \right],\\
&n = {1},{2},...,{N},~l = {0},{1},...,{L-1}.
\end{split}
\label{data_entry}
\end{equation}
For notational simplicity, we drop the symbol index $k$ and vectorize (\ref{data_entry}) to
\begin{equation}
{{\mathbf{y}}_{l}}=\sqrt{{{E}_{s}}}b{{s}_{l}}{{\mathbf{H}}^{T}}\mathbf{w}+{{\mathbf{i}}_{l}}+{{\text{\boldmath$\omega$} }_{l}}, \, \,\, l=0,1,...,L-1, \label{ynl}
\end{equation} 
where $\mathbf{H}\in {{\mathbb{C}}^{M\times N}}$ is the MIMO-formed channel matrix with elements $\mathbf{H}[i,j] = h_{i,j}$, ${{\text{\boldmath$\omega$} }_{l}}\in {{\mathbb{C}}^{N}}$ is a complex white Gaussian noise vector with autocorrelation matrix ${{\mathbf{R}}_{\omega }}= \sigma _{\omega }^{2}{{\mathbf{I}}_{N}}$ and ${{\mathbf{i}}_{l}}\in {{\mathbb{C}}^{N}}$ models comprehensively environmental disturbance of any other form. 

%----------------

%The MIMO system model is depicted in Fig. 1. 
We denote ${{\mathbf{o}}_{l}}={{\mathbf{i}}_{l}}+{{\text{\boldmath$\omega$} }_{l}}$ as the signal occupancy in the band sensed at carrier frequency $f_c$. Considering that (\ref{ynl}) is both space and time representation of collected values, we can organize the space-time data using the code vector ${{\mathbf{s}}_{L\times 1}}$ as
\begin{equation}
\begin{split}
{{\mathbf{y}}_{NL\times 1}}&=\sqrt{{{E}_{s}}}b\left( \mathbf{s}\otimes {{\mathbf{H}}^{T}}\mathbf{w} \right)+{{\mathbf{o}}_{NL\times 1}} \\
&=\sqrt{{{E}_{s}}}b\mathbf{g}+{{\mathbf{o}}_{NL\times 1}},
\end{split}
\end{equation}
where
\begin{equation}
\mathbf{g}=\mathbf{s}\otimes \left( {{\mathbf{H}}^{T}}\mathbf{w} \right)\in {{\mathbb{C}}^{NL\times 1}}, \label{gvec}
\end{equation}
and the “\emph{space-time occupancy autocorrelation}” matrix is
\begin{equation}
{{\mathbf{O}}_{{{f}_{c}}}}\triangleq E\left\{ \left( {{\mathbf{o}}_{NL\times 1}} \right){{\left( {{\mathbf{o}}_{NL\times 1}} \right)}^{H}} \right\}\in {{\mathbb{C}}^{NL\times NL}}. \label{Ofc}
\end{equation}

\section{Joint Space-Time Waveform Shaping}
\subsection{Waveform Design Problem}
%The SINR as a function of code sequence and beam weight vector can be written as
% \begin{equation}
% \begin{split}
% SINR\left( \mathbf{s},\mathbf{w} \right)&\triangleq {{E}_{s}}\frac{E\left\{ {{\left| {{\mathbf{g}}^{H}}\mathbf{g} \right|}^{2}} \right\}}{E\left\{ {{\left| {{\mathbf{g}}^{H}}{{\mathbf{o}}_{NL\times 1}} \right|}^{2}} \right\}} \\
% &={{E}_{s}}\frac{E\left\{ {{\mathbf{g}}^{H}}\mathbf{g}{{\mathbf{g}}^{H}}\mathbf{g} \right\}}{E\left\{ {{\mathbf{g}}^{H}}{{\mathbf{o}}_{NL\times 1}}\mathbf{o}_{NL\times 1}^{H}\mathbf{g} \right\}}.
% \end{split}
% \end{equation}
The space-time maximum SINR receiver filter is given as $\mathbf{O}_{{{f}_{c}}}^{-1}\mathbf{g}$. The output SINR of the maximum SINR space-time receiver filter can be written as a function of the code sequence and the transmit beam weight vector as follows
\begin{equation}
\begin{split}
{\rm{SINR}}&\triangleq {{E}_{s}}\frac{E\left\{ {{\left| {{\mathbf{g}}^{H}}{{ \mathbf{O}_{{{f}_{c}}}^{-1} }}\mathbf{g} \right|}^{2}} \right\}}{E\left\{ {{\left| {{\mathbf{g}}^{H}}{{ \mathbf{O}_{{{f}_{c}}}^{-1} }}{{\mathbf{o}}_{NL\times 1}} \right|}^{2}} \right\}} \\
&=\frac{{{E}_{T}}}{M}\frac{E\left\{ {{\mathbf{g}}^{H}}{{ \mathbf{O}_{{{f}_{c}}}^{-1} }}\mathbf{g}{{\mathbf{g}}^{H}}\mathbf{O}_{{{f}_{c}}}^{-1}\mathbf{g} \right\}}{E\left\{ {{\mathbf{g}}^{H}}{{ \mathbf{O}_{{{f}_{c}}}^{-1} }}{{\mathbf{o}}_{NL\times 1}}\mathbf{o}_{NL\times 1}^{H}\mathbf{O}_{{{f}_{c}}}^{-1}\mathbf{g} \right\}},\label{SINR}
\end{split}
\end{equation}
where $\mathbf{g}=\mathbf{s}\otimes \left( {{\mathbf{H}}^{T}}\mathbf{w} \right)\in {{\mathbb{C}}^{NL\times 1}}$ shows the dependency on code sequence $\mathbf{s}$ and transmit beam weigh vector $\mathbf{w}$. Therefore, it is of interest to investigate waveform designs for a locally sensed space-time occupancy autocorrelation matrix ${{\mathbf{{O}}}_{{{f}_{c}}}}$.

The purpose of space-time waveform shaping implemented by the new wireless MIMO link for spectrum sharing is to optimally maintain a predefined pre-detection SINR value $\gamma $ (minimum required SINR for successful operation) while minimizing the total transmit energy per symbol by optimally selecting the beam weight vector $\mathbf{w}$, pulse code sequence $\mathbf{s}$, and transmit energy ${{E}_{T}}$. In this way, the transceiver optimally shapes the transmitted waveform and reception jointly in time and space to coexist in a congested frequency band in two manners: (a) optimally avoid sensed interference and disturbance by adaptively reshaping the waveform and (b) optimally protect the neighbor links by minimizing the transmitted power in the band.

%\vspace{-0.2cm}
%\subsection{Problem Formulation}
Considering SINR in (\ref{SINR}), the wavefom optimization problem is given by $\min \text{ }{{E}_{T}}$ subject to the constraint that ${\rm{SINR}}\left( {{E}_{T}} \right)\ge \gamma$. Since ${\rm{SINR}}\left( {{E}_{T}} \right)$ is linearly proportional to $E_T$, this can be written as $\min \text{ }{{E}_{T}}$ s.t. ${\rm SINR}\left( {{E}_{T}} \right)= \gamma$. Therefore, we rewrite (\ref{SINR}) as
\begin{equation}
\begin{split}
{{E}_{T}}\left( \mathbf{s},\mathbf{w} \right)&=\gamma M\frac{E\left\{ {{\mathbf{g}}^{H}}{{\mathbf{O}_{{{f}_{c}}}^{-1}}}{{\mathbf{o}}_{NL\times 1}}\mathbf{o}_{NL\times 1}^{H}\mathbf{O}_{{{f}_{c}}}^{-1}\mathbf{g} \right\}}{E\left\{ {{\mathbf{g}}^{H}}{{ \mathbf{O}_{{{f}_{c}}}^{-1}}}\mathbf{g}{{\mathbf{g}}^{H}}\mathbf{O}_{{{f}_{c}}}^{-1}\mathbf{g} \right\}} \\
&=\frac{\gamma M}{{{\mathbf{g}}^{H}}\mathbf{O}_{{{f}_{c}}}^{-1}\mathbf{g}}. \label{ETsw}
\end{split}
\end{equation}

\vspace{-0.2cm}
Further, we insert an upper bound on the peak transmit energy, i.e., ${{E}_{T}}\le {{E}_{T,\max }}$, where ${{E}_{T,\max }}$ is the peak allowable total transmit energy per symbol. When we cap the transmit energy, there may be circumstances that the transceiver is not able to maintain the predefined pre-detection SINR value of $\gamma $. In that scenario, the transceiver refrains from transmitting and instead maintains co-channel non-cooperative links that occupy the band, meaning ${{E}_{T}}=0$. To differentiate the two scenarios and formulate the problem, we define the parameter $\gamma_{\max}$ as the maximum possible attainable SINR given a channel matrix $\mathbf{H}$ and an occupancy autocorrelation matrix ${{\mathbf{O}}_{{{f}_{c}}}}$. 
To find the maximum attainable SINR of a given channel matrix and occupancy autocorrelation matrix, we set ${{E}_{T}}={{E}_{T,\max }}$ and we have
\begin{equation}
    \gamma_{\max}=\frac{{{E}_{T,\max }}}{M}\left( \max {{\mathbf{g}}^{H}}\mathbf{O}_{{{f}_{c}}}^{-1}\mathbf{g} \right),\label{SINRmax}
\end{equation}
where the maximum in (\ref{SINRmax}) is achieved when $\mathbf{g}={{\mathbf{q}}_{\max }}$ the maximum-eigenvalue eigenvector of the inverse of joint space-time occupancy autocorrelation matrix $\mathbf{O}_{{{f}_{c}}}^{-1}$. Defining ${{\lambda }_{\max }}=\mathbf{q}_{\max }^{H}\mathbf{O}_{{{f}_{c}}}^{-1}{{\mathbf{q}}_{\max }}$ as the maximum eigenvalue of $\mathbf{O}_{{{f}_{c}}}^{-1}$, we have the maximum attainable SINR as $\gamma_{\max}={{E}_{T,\max }}{{\lambda }_{\max }}/M$. 

%\subsection{Optimization Solution}
By replacing $\mathbf{g}$ from (\ref{gvec}) in (\ref{ETsw}), we re-write the waveform design problem as
%\begin{equation}
%\begin{split}
%  & E_{T}^{opt}= \underset{\mathbf{s},\mathbf{w}}{\mathop{\min }}\,\frac{\gamma M}{{{[\mathbf{s}\otimes \left( {{\mathbf{H}}^{T}}\mathbf{w} \right)]}^{H}}\mathbf{O}_{{{f}_{c}}}^{-1}[\mathbf{s}\otimes \left( {{\mathbf{H}}^{T}}\mathbf{w} \right)]}\\ 
 %& \text{s}\text{.t}\text{.} \\ 
 %& \text{~~~~~~1)~~~} E_{T}^{opt}\le E_{T,max}\\ 
 %& \text{~~~~~~2)~~~}\mathbf{s}\in {{\left\{ \pm 1/\sqrt{L},\pm j/\sqrt{L} \right\}}^{L}} \\ 
 %& \text{~~~~~~3)~~~}\mathbf{w}\in %{{\mathbb{C}}^{M}},\text{ }{{\left\| \mathbf{w} %\right\|}^{2}}=M \\ 
%\end{split}
%\end{equation}

\begin{gather}
    E_{T}^{opt}= \underset{\mathbf{s},\mathbf{w}}{\mathop{\min }}\,\frac{\gamma M}{{{[\mathbf{s}\otimes \left( {{\mathbf{H}}^{T}}\mathbf{w} \right)]}^{H}}\mathbf{O}_{{{f}_{c}}}^{-1}[\mathbf{s}\otimes \left( {{\mathbf{H}}^{T}}\mathbf{w} \right)]}\\
    \begin{aligned}
        \text{s.t.} \quad   &E_{T}^{opt}\le E_{T,max}           \nonumber\\
                            &\mathbf{s}\in {{\left\{ \pm 1/\sqrt{L},\pm j/\sqrt{L} \right\}}^{L}}            \nonumber\\    
                            &\mathbf{w}\in {{\mathbb{C}}^{M}},\text{ }{{\left\| \mathbf{w} \right\|}^{2}}=M
                            %\label{eqn:nonnegative}
    \end{aligned}
\end{gather}

\subsection{Joint Space-Time Optimization}
The optimization problem succeeds in optimal values to approximating closely the vector $\mathbf{g}$ to the eigenvector ${{\mathbf{q}}_{\max }}$, i.e., we try to solve
\begin{equation}
\left( {{\mathbf{s}}^{opt}},{{\mathbf{w}}^{opt}} \right)=\underset{\mathbf{s},\mathbf{w}}{\mathop{\arg \min }}\,{{\left\| {{\mathbf{q}}_{\max }}-\left( \mathbf{s}\otimes {{\mathbf{H}}^{T}}\mathbf{w} \right) \right\|}^{2}}.\label{argmin}
\end{equation}
Therefore, to achieve the optimal space-time waveform shaping, the total transmit energy per symbol is
\begin{equation}
E_{T}^{opt}=\left\{ \begin{matrix}
   \gamma M/{{\lambda }_{\max }}\text{~~~~~~~}\gamma \le {{\gamma }_{\max }}  \\
   0\text{~~~~~~~~~~~~~~~~~}\gamma >{{\gamma }_{\max }}  \\
\end{matrix} \right.
.
\label{ETopt}
\end{equation}
The upper expression in (\ref{ETopt}) guarantees $E_{T}^{opt}\le {{E}_{T,\max }}$ and implements space-time waveform shaping to maintain minimum required SINR, $\gamma$, for spectrum sharing with co-channel operating links. According to the lower expression in (\ref{ETopt}), we refrain from transmitting and leave the bandin favor of co-channel nodes because even with optimal waveform shaping, it is not possible to maintain $\gamma$ for successful operation (other than exceeding ${{E}_{T,\max }}$), hence leaving the band in favor of co-channel nodes.  We can prove that a closed-form expression of ${{\mathbf{w}}^{opt}}$ for any fixed code vector $\mathbf{s}$ can be found. We consider the gradient of the objective function in (\ref{argmin}) with respect to ${{\mathbf{w}}^{H}}$. We expand the ${{l}_{2}}$-norm and apply the Hermitian operator to all components inside the first parenthesis,
\begin{equation}
\begin{split}
  & {{\nabla }_{{{\mathbf{w}}^{H}}}}{{\left\| {{\mathbf{q}}_{\max }}-\left( \mathbf{s}\otimes {{\mathbf{H}}^{T}} \right)\mathbf{w} \right\|}^{2}}= \\ 
  & {{\nabla }_{{{\mathbf{w}}^{H}}}}\left[ {{\left( {{\mathbf{q}}_{\max }}-\left( \mathbf{s}\otimes {{\mathbf{H}}^{T}} \right)\mathbf{w} \right)}^{H}}\left( {{\mathbf{q}}_{\max }}-\left( \mathbf{s}\otimes {{\mathbf{H}}^{T}} \right)\mathbf{w} \right) \right]= \\ 
 & {{\nabla }_{{{\mathbf{w}}^{H}}}}\left[ \left( \mathbf{q}_{\max }^{H}-{{\mathbf{w}}^{H}}\left( {{\mathbf{s}}^{H}}\otimes {{\mathbf{H}}^{*}} \right) \right)\left( {{\mathbf{q}}_{\max }}-\left( \mathbf{s}\otimes {{\mathbf{H}}^{T}} \right)\mathbf{w} \right) \right]. \\ 
\end{split}
\end{equation}
We set the gradient equal to $\mathbf{0}\in {{\mathbb{C}}^{M}}$ and calculate
\begin{equation}
-\left( {{\mathbf{s}}^{H}}\otimes {{\mathbf{H}}^{*}} \right){{\mathbf{q}}_{\max }}+\left( {{\mathbf{s}}^{H}}\otimes {{\mathbf{H}}^{*}} \right)\left( \mathbf{s}\otimes {{\mathbf{H}}^{T}} \right)\mathbf{w}={{\mathbf{0}}_{M\times 1}}.\label{equal}
\end{equation}
We solve (\ref{equal}) to obtain
\begin{equation}
{{\mathbf{w}}^{opt}}=inv\left[ \left( {{\mathbf{s}}^{H}}\otimes {{\mathbf{H}}^{*}} \right)\left( \mathbf{s}\otimes {{\mathbf{H}}^{T}} \right) \right]\left( {{\mathbf{s}}^{H}}\otimes {{\mathbf{H}}^{*}} \right){{\mathbf{q}}_{\max }}\label{wopt}
\end{equation}
where $\left( {{\mathbf{s}}^{H}}\otimes {{\mathbf{H}}^{*}} \right)\left( \mathbf{s}\otimes {{\mathbf{H}}^{T}} \right)$ can be further simplified using properties of Kronecker product as
\begin{equation}
\begin{split}
   {{\mathbf{w}}^{opt}}&=inv\left[ \left( {{\mathbf{s}}^{H}}\mathbf{s}\otimes {{\mathbf{H}}^{*}}{{\mathbf{H}}^{T}} \right) \right]\left( {{\mathbf{s}}^{H}}\otimes {{\mathbf{H}}^{*}} \right){{\mathbf{q}}_{\max }} \\ 
 & =inv\left[ \left( 1\otimes {{\mathbf{H}}^{*}}{{\mathbf{H}}^{T}} \right) \right]\left( {{\mathbf{s}}^{H}}\otimes {{\mathbf{H}}^{*}} \right){{\mathbf{q}}_{\max }} \\ 
 & ={{\left( {{\mathbf{H}}^{*}}{{\mathbf{H}}^{T}} \right)}^{-1}}\left( {{\mathbf{s}}^{H}}\otimes {{\mathbf{H}}^{*}} \right){{\mathbf{q}}_{\max }} \\ 
\end{split}\label{wopt2}
\end{equation}
where we consider that ${{\mathbf{s}}^{H}}\mathbf{s}=1$ and ${{\mathbf{H}}^{*}}{{\mathbf{H}}^{T}}$ is invertible if $rank\left( \mathbf{H} \right)>M$. Inserting (\ref{wopt2}) in (\ref{gvec}), we calculate ${{\mathbf{g}}^{opt}}$ as
\begin{equation}
\begin{split}
 {{\mathbf{g}}^{opt}}&=\left( \mathbf{s}\otimes {{\mathbf{H}}^{T}} \right){{\left( {{\mathbf{H}}^{*}}{{\mathbf{H}}^{T}} \right)}^{-1}}\left( {{\mathbf{s}}^{H}}\otimes {{\mathbf{H}}^{*}} \right){{\mathbf{q}}_{\max }} \\ 
 & =\left( \mathbf{s}\otimes \left[ {{\mathbf{H}}^{T}}{{\left( {{\mathbf{H}}^{*}}{{\mathbf{H}}^{T}} \right)}^{-1}} \right] \right)\left( {{\mathbf{s}}^{H}}\otimes {{\mathbf{H}}^{*}} \right){{\mathbf{q}}_{\max }} \\ 
 & =\left( \mathbf{s}{{\mathbf{s}}^{H}}\otimes \left[ {{\mathbf{H}}^{T}}{{\left( {{\mathbf{H}}^{*}}{{\mathbf{H}}^{T}} \right)}^{-1}}{{\mathbf{H}}^{*}} \right] \right){{\mathbf{q}}_{\max }} \\ 
 & =\left( \mathbf{s}{{\mathbf{s}}^{H}}\otimes {{\mathbf{I}}_{N\times N}} \right){{\mathbf{q}}_{\max }}. \\ 
\end{split}\label{gopt}
\end{equation}

Inserting now (\ref{gopt}) in (\ref{argmin}), we can find the jointly optimal code vector ${{\mathbf{s}}^{opt}}$ with a simple search
\begin{equation}
\begin{split}
   {{\mathbf{s}}^{opt}}&=\underset{\mathbf{s}}{\mathop{\arg \min }}\,{{\left\| {{\mathbf{q}}_{\max }}-\left( \mathbf{s}{{\mathbf{s}}^{H}}\otimes {{\mathbf{I}}_{N\times N}} \right){{\mathbf{q}}_{\max }} \right\|}^{2}} \\ 
 & =\underset{\mathbf{s}}{\mathop{\arg \min }}\,{{\left\| \left\{ {{\mathbf{I}}_{NL\times NL}}-\left( \mathbf{s}{{\mathbf{s}}^{H}}\otimes {{\mathbf{I}}_{N\times N}} \right) \right\}{{\mathbf{q}}_{\max }} \right\|}^{2}} \\  
\end{split}\label{sopt}
\end{equation}
where $\mathbf{q}_{\max }$ is the $NL\times 1$ eigenvector. Algorithm (\ref{algorithm1}) summarizes the proposed joint space-time waveform shaping optimization algorithm. %is summarized for easy reference in the following.

\begin{algorithm}
    \caption{Joint space-time waveform shaping}
    \label{algorithm1}
  \begin{algorithmic}[1]
    \INPUT Pulse-filtered interference-plus-noise received samples; estimated channel matrix $\mathbf{H}\in {{\mathbb{C}}^{M\times N}}$; SINR threshold $\gamma $; peak transmit energy ${{E}_{T,\max }}$.
    \STATE Calculate (estimate) space-time occupancy autocorrelation matrix ${{\mathbf{O}}_{{{f}_{c}}}}$ in (\ref{Ofc}).
    \STATE Calculate maximum-eigenvalue eigenvector of ${{\mathbf{O}}_{{{f}_{c}}}^{-1}}$, $\mathbf{q}_{max}$.
    \STATE Find optimum code $\mathbf{s}^{opt}$ by discrete search over (\ref{sopt})
    \STATE Find jointly optimal beam weight vector $\mathbf{w}^{opt}$ by inserting $\mathbf{s}^{opt}$ in (\ref{wopt2}).
    \STATE Find $\gamma_{max}$ by inserting $\mathbf{w}^{opt}$ and $\mathbf{s}^{opt}$ in (\ref{SINRmax}).
    \STATE Compare the predefined SINR threshold $\gamma $ with $\gamma_{max}$ and adjust $E_{T}^{opt}$.
    \OUTPUT $E_{T}^{opt}$, $\mathbf{s}^{opt}$, $\mathbf{w}^{opt}$
  \end{algorithmic}
\end{algorithm}

\section{Numerical Studies}

%In this section, we present simulation results to collaborate analytical derivations. 
We consider a  $4\times 4$ MIMO system operating in  light and heavily congested frequency bands. % We present the performance of optimal waveform shaping highlighting the improvements. 
We consider an operating frequency of ${{f}_{c}}=900 $~MHz, 64-QAM modulation and code sequence lengths of $L=4$ and 8. The bandwidth of the transmitted signal is $BW=\left( 1+\alpha  \right)/{{T}_{c}}$ and the data rate equivalent is $R={{\log }_{2}}\left( Q \right)/L{{T}_{c}}$ bits/second. As explained in section II.a, ${{E}_{T}}$ is the total transmit energy per symbol. For ease of presentation and clarity, we consider ${{E}_{T}}$ as the total transmit energy per symbol divided by ${{N}_{0}}$. Therefore, the values presented as energy are multiples of ${{N}_{0}}$. For example, we consider the predefined value of ${{E}_{T,\max }}=20$ i.e., the peak allowable total transmit energy per symbol is 20x  ${{N}_{0}}$. Providing a realistic value in Joules, we assume thermal noise is at 300 Kelvin and the bandwidth to be 100 MHz. The noise power is $N_0=K_bBT$ where $K_b=1.38064852\times10^{-23} J.K^{-1}$ is the Boltzmann constant. Therefore, we have $N_0=0.414\times10^{-12} W$ and ${{E}_{T,\max }}=4.97\times10^{-19} J$ if we consider a symbol duration of 0.6 ns. For setting up the pre-detection SINR threshold $\gamma$ of the new MIMO wireless data link, we consider %the threshold as when 64-QAM modulation 
the BER performance to be almost ${{10}^{-5}}$.  This is equivalent to $\gamma =18$ dB for 64-QAM modulation, which is fixed in our simulation results. To model the occupied band, we consider spread-spectrum and narrowband signals operating at the same frequency band with our MIMO data link. In particular, these signals are described by 
\begin{equation}
{{i}_{{{\rm NB}}}}\left( t \right)=\sum\limits_{k}^{{}}{\sqrt{{{E}_{i}}}{b}_{{{\rm NB}}}\left[ k \right]p\left( t-k{{T}_{s}} \right)\mathbf{H}_{{{\rm NB}}}^{T}{{\mathbf{w}}_{{{\rm NB}}}}},
\end{equation}
for narrowband signal occupying the band with ${{\mathbf{w}}_{{{\rm NB}}}}$ transmit beam weight vector over ${{M}_{{{\rm NB}}}}$ transmit antennas, ${b}_{{\rm NB}}\left[ k \right]\in $64-QAM alphabet, and bandwidth $\left( 1+\alpha  \right)/{{T}_{s}}$. The spread-spectrum signal occupying the band is
\begin{equation}
{{i}_{{{\rm SS}}}}\left( t \right)=\sum\limits_{k}^{{}}{\sqrt{{{E}_{i}}}{b}_{{{\rm SS}}}\left[ k \right]{\bar s}\left( t-k{{T}_{s}} \right)\mathbf{H}_{{{\rm SS}}}^{T}{{\mathbf{w}}_{{{\rm SS}}}}},
\end{equation}
\begin{equation}
{\bar s}\left( t \right)=\sum\limits_{l=0}^{L-1}{{{{\bar{s}}}_{l}}p\left( t-l{{T}_{c}} \right),}
\end{equation}
with ${{\bar{s}}_{l}}\in \left\{ \pm 1/\sqrt{L},\pm j/\sqrt{L} \right\}$, ${{\mathbf{w}}_{{{\rm SS}}}}$ transmit beam weight vector over ${{M}_{{{\rm SS}}}}$ transmit antennas, ${b}_{{{\rm SS}}}\left[ k \right]\in $64-QAM alphabet, and bandwidth $L \left \{\left( 1+\alpha  \right)/{{T}_{s}}\right \}$.

\begin{figure}[t]
\centering
\subfloat[]{\includegraphics[width=88mm,scale=0.9]{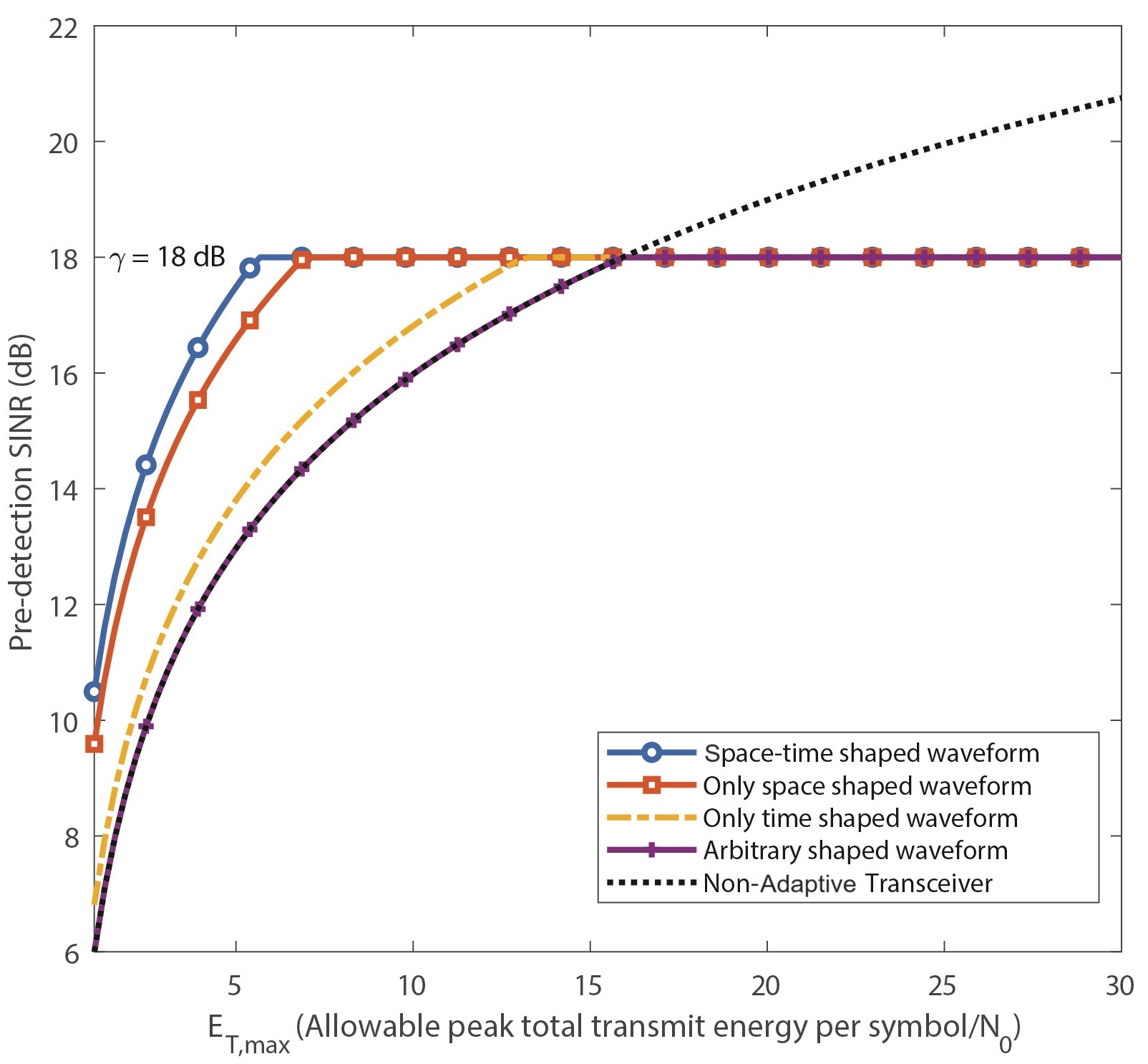}}\\
\subfloat[]{\includegraphics[width=88mm,scale=0.9]{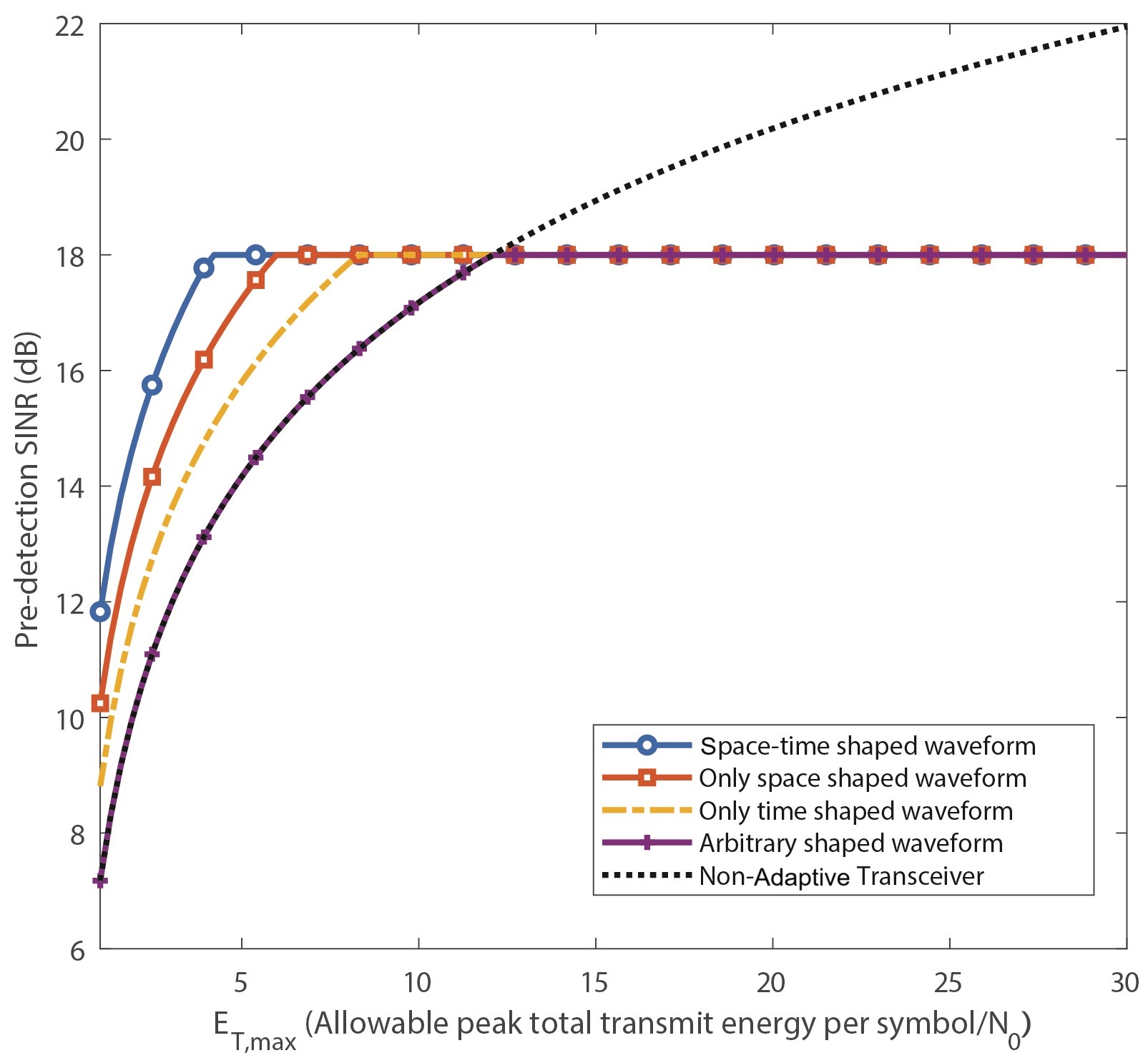}}
\caption{Pre-detection SINR of $4\times 4$ optimal waveform shaping MIMO system ($\gamma =18$ dB): (a) $L=4$, (b) $L=8$.}
\end{figure}

\begin{figure}[t]
\centering
\subfloat[]{\includegraphics[width=85mm,scale=0.9]{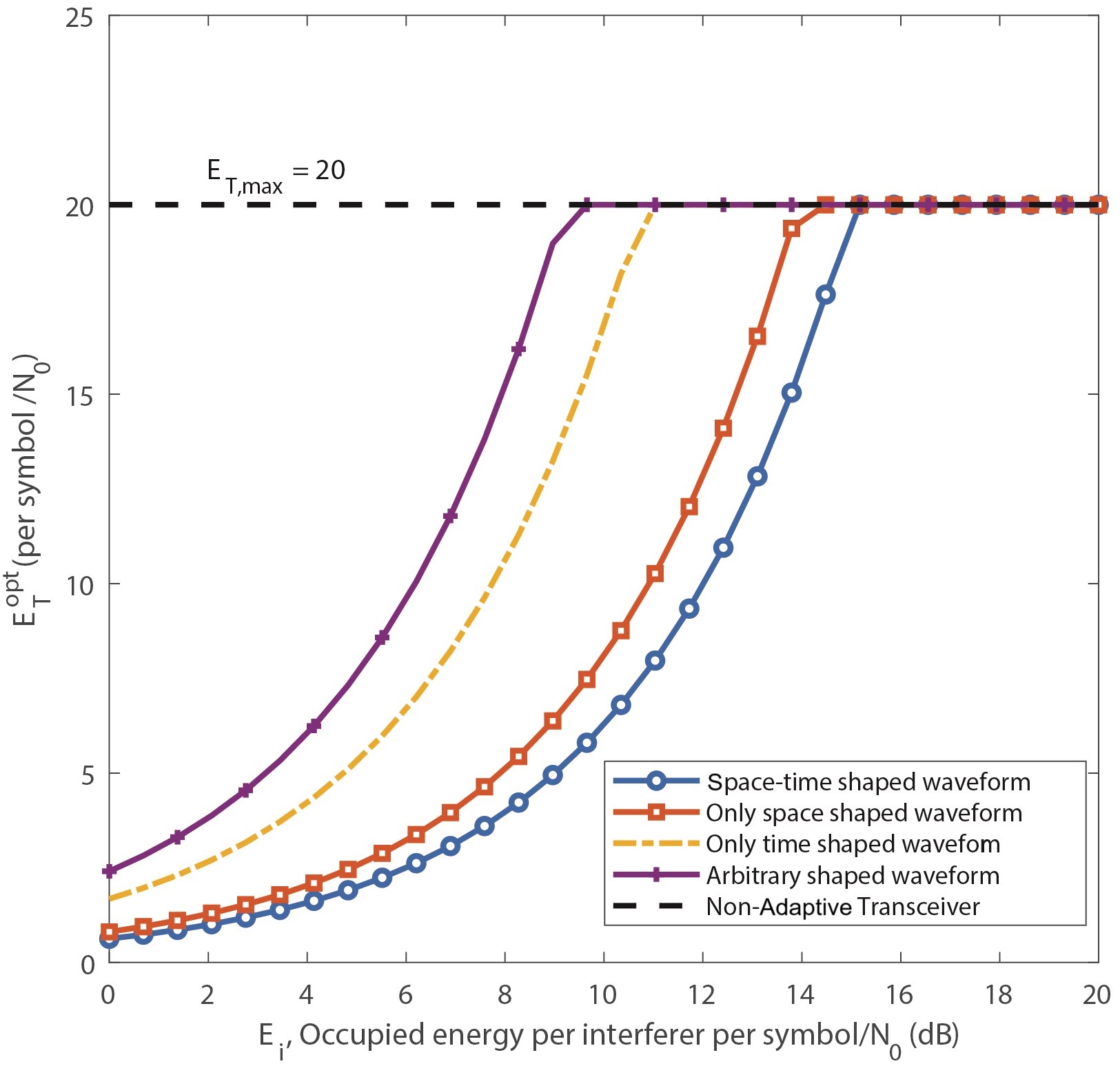}}\\
\subfloat[]{\includegraphics[width=85mm,scale=0.9]{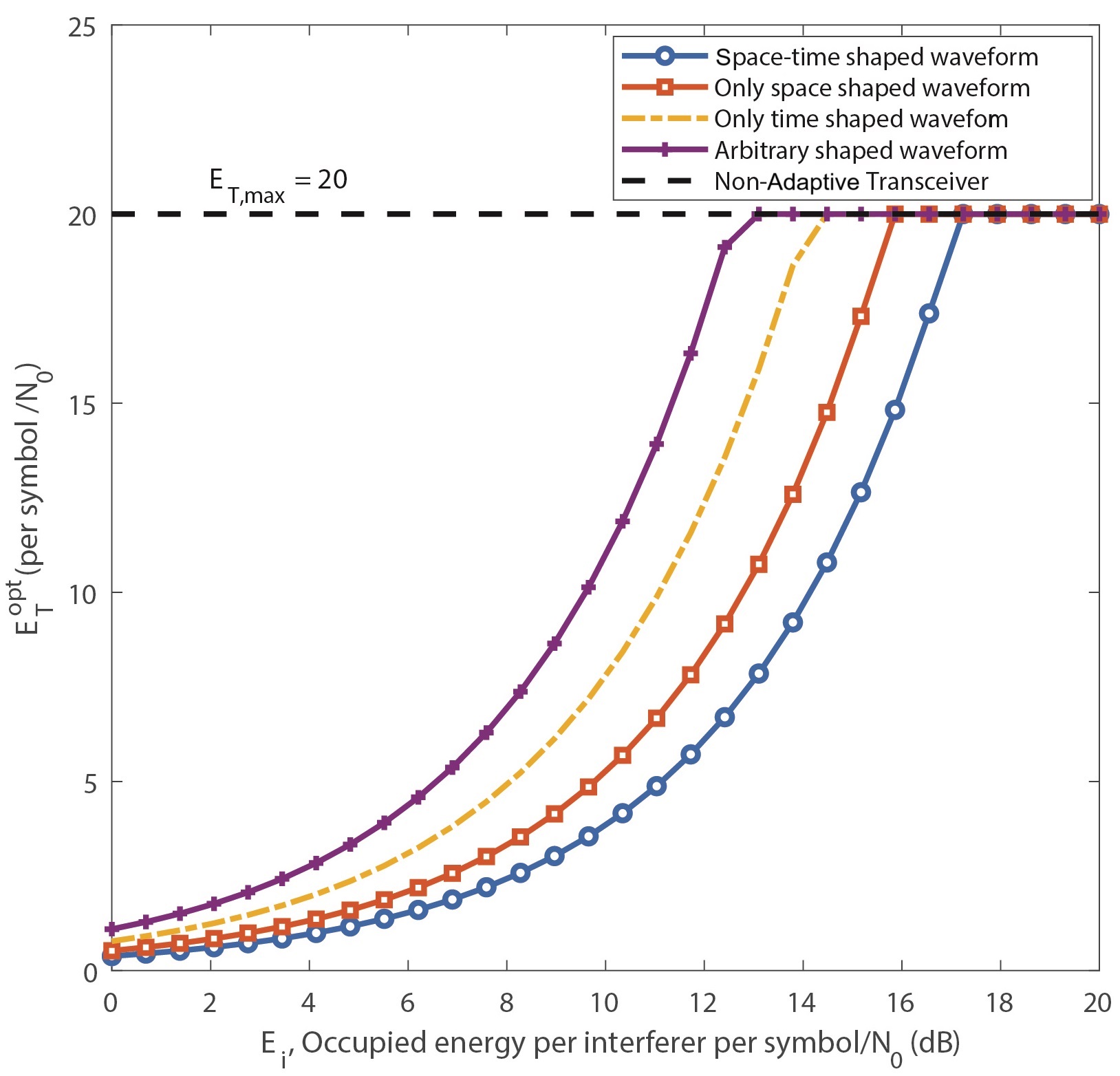}}
\caption{Optimal total transmit energy per symbol over ${{N}_{0}}$ ($4\times 4$, $\gamma =18$ dB): (a) $L=4$, (b) $L=8$.}
\end{figure}

In Fig. 1a, we study the MIMO system pre-detection SINR and plot it according to the predefined allowable peak total transmit energy i.e., ${{E}_{T,\max }}$ assuming $\gamma =18$ dB and $L=4$. Specifically, we change the value of ${{E}_{T,\max }}$ from 1 to 30 times ${{N}_{0}}$. The band is occupied with narrowband and spread-spectrum signals with ${{M}_{{\rm NB}}}={{M}_{{\rm SS}}}=4$ and ${{E}_{i}}=10$ dB. Particularly, we study the MIMO system's SINR for the cases of: 1) Joint space-time shaped waveform described in section III; 2) Space-only shaped waveform in which we consider a fixed arbitrary code sequence $\mathbf{s}$, while ${{\mathbf{w}}^{opt}}$ and $E_{T}^{opt}$ are used; 3) Time-only shaped waveform in which we consider a fixed arbitrary transmit beam weight vector $\mathbf{w}$, while ${{\mathbf{s}}^{opt}}$ and $E_{T}^{opt}$ are used; 4) Arbitrary shaped waveform in which both $\mathbf{s}$ and $\mathbf{w}$ are fixed and arbitrary, while $E_{T}^{opt}$ is used; and 5) a non-adaptive transceiver as a benchmark. We observe that the joint space-time shaped waveform achieves the highest gain and then it is followed by the space-only shaped waveform, time-only shaped waveform, and finally an arbitrary shaped waveform. Specifically, assuming ${{E}_{T,\max }}=5.4$, we have the SINR values of 17.81, 16.90, 14.13, and 13.29 respectively for the above-mentioned cases. It means that a gain of 4.52 dB is achieved for the joint space-time shaped waveform in compared to arbitrary case. This gain changes to 3.61 and 0.84 dB for space-only and time-only shaped waveforms respectively. The performance of the non-adaptive transceiver and arbitrary shaped waveform are the same up until they achieve the predefined SINR threshold of $\gamma =18$ dB. In comparison to the non-adaptive transceiver, arbitrary shaped waveform avoids consuming excessive energy transmitted in band for the SINR regions above 18 dB and adjusts the total transmit energy to only maintain the SINR threshold. At the same time, we observe that the proposed joint space-time shaped waveform maintains the SINR threshold earlier with much smaller values of transmit energy followed by the space-only shaped waveform and then the time-only shaped waveform. Particularly, the joint space-time shaped waveform maintains the SINR threshold at ${{E}_{T,\max }}=5.68$, space-only shaped waveform at ${{E}_{T,\max }}=6.85$, time-only shaped waveform at ${{E}_{T,\max }}=13.01$, and arbitrary shaped waveform at ${{E}_{T,\max }}=15.64$. Further, we increase the code length parameter $L$ to 8 and observe the changes in Fig. 1b. In general, the same pattern is repeated for the cases except for the fact that in general all the gains are improved. Particularly, this improvement is much more pronounced for the time-only shaped waveform. Assuming the same value ${{E}_{T,\max }}=5.4$, the time-only shaped waveform achieves the SINR value of 16.13 dB while the arbitrary shaped waveform achieves 14.49 dB, providing a 1.64 dB gain which is higher than the same case for $L=4$.

In Fig. 2a, we plot $E_{T}^{opt}$ i.e., the optimal total transmit energy per symbol over ${{N}_{0}}$ (a parameter of waveform shaping) while we create a dynamic occupied band by changing/increasing ${{E}_{i}}$, the occupied energy per interferer per symbol over ${{N}_{0}}$ from 0 dB to 20 dB. Assuming $\gamma =18$ dB and $L=4$, we plot the optimal energy for all aforementioned cases of waveform shaping and a non-adaptive transceiver. We observe that the adaptive waveform shaping MIMO link easily outperforms the non-adaptive transceiver by putting less total transmit energy in the band to maintain the SINR threshold. In this way, it also protects other users that are already operating in the band by minimizing the amount of energy disturbance. As expected, the order of performance is the same as before with the proposed joint space-time shaped waveform providing the highest gain followed by space-only and time-only shaped waveforms. For instance, at a value of ${{E}_{i}}=8.27$ dB, the optimal total transmit energy is calculated as $E_{T}^{opt}=4.22$ for the joint space-time shaped waveform. This increases to $E_{T}^{opt}=5.43$, 11.29, and 16.19 for space-only, time-only, and arbitrary shaped waveforms,  respectively. A non-adaptive transceiver typically transmits at the maximum allowable energy, i.e., ${{E}_{T,\max }}=20$ always occupying the band with maximum energy. The arbitrary shaped waveform only is forced to transmit at maximum energy when it cannot maintain the SINR threshold, i.e., ${{E}_{i}}\ge 9.65$ dB. This value changes to ${{E}_{i}}\ge 11.03$ dB for time-only, ${{E}_{i}}\ge 14.48$ dB for space-only, and ${{E}_{i}}\ge 15.17$ dB for joint space-time shaped waveform manifesting its higher resistance to share the spectrum in the heavily occupied band. In Fig. 2b, we depict the results for $L=8$. An overall performance improvement is observed for all the cases while a pronounced improved gain is achieved for time-only shaped waveform, as expected. % in trade off for doubling the bandwidth or halving the data rate.
Ultimately, the joint space-time shaped waveform resists up to ${{E}_{i}}=17.24$ dB and demonstrated gains of 2.07 dB when compared to the same case for $L=4$.

\section{Conclusion}
%We considered the task of introducing a new MIMO wireless data link over a given occupied frequency band. 
We proposed and evaluated through simulations, in light and heavily congested band scenarios,
%discussed the modeling and optimization problem details of a MIMO transceiver which optimally shapes the transmitted waveform and reception jointly in time and space to coexist in a utilized frequency band in two manners: 
spectrum sharing by a new MIMO wireless data link that (a) optimally avoids sensed interference in the joint space-time domain, and (b) protects existing links by minimizing its transmitted power in the band. Link adaptation is based on the estimated channel coefficients and sensed occupancy autocorrelation matrix. In particular, the MIMO transmit beam weight vector and time-domain pulse code sequence are jointly optimized to minimize the transmit energy per bit per antenna, while maintaining a pre-defined SINR threshold at the receiver. %We formulated the design of optimal waveform shaping MIMO transceiver and extensive numerical results are provided to demonstrate derived algorithmic solutions under spread-spectrum/non-spread-spectrum interferences, in light and dense interference scenarios. 
%The transceiver dynamically optimizes beam weight vector and the pulse code sequence to shape the waveform and transmitted energy. 
We show that the proposed autonomously reconfigurable 4x4 MIMO link outperforms a non-adaptive transceiver and other forms of waveform shaping in terms of the pre-detection SINR performance and the capability to protect ongoing non-cooperative links by not occupying the band with redundant transmissions. 
We observe that the proposed joint space-time waveform shaping is capable of maintaining the SINR threshold required for the new MIMO data link with much lower transmit energy when compared to space-only shaped and time-only shaped waveforms. Furthermore, by increasing the number of coded repeats of pulses, higher gains are demonstrated by the proposed space-time shaped waveform.

\bibliographystyle{IEEEtran}
\bibliography{bare_jrnl}

\begin{thebibliography}{10}
\providecommand{\url}[1]{#1}
\csname url@samestyle\endcsname
\providecommand{\newblock}{\relax}
\providecommand{\bibinfo}[2]{#2}
\providecommand{\BIBentrySTDinterwordspacing}{\spaceskip=0pt\relax}
\providecommand{\BIBentryALTinterwordstretchfactor}{4}
\providecommand{\BIBentryALTinterwordspacing}{\spaceskip=\fontdimen2\font plus
\BIBentryALTinterwordstretchfactor\fontdimen3\font minus \fontdimen4\font\relax}
\providecommand{\BIBforeignlanguage}[2]{{%
\expandafter\ifx\csname l@#1\endcsname\relax
\typeout{** WARNING: IEEEtran.bst: No hyphenation pattern has been}%
\typeout{** loaded for the language `#1'. Using the pattern for}%
\typeout{** the default language instead.}%
\else
\language=\csname l@#1\endcsname
\fi
#2}}
\providecommand{\BIBdecl}{\relax}
\BIBdecl

\bibitem{ref1}
G.~Sklivanitis, A.~Gannon, S.~N. Batalama, and D.~A. Pados, ``Addressing next-generation wireless challenges with commercial software-defined radio platforms,'' \emph{IEEE Commun. Mag.}, vol.~54, no.~1, pp. 59--67, Jan. 2016.

\bibitem{ref2}
Y.~Cao, T.~Jiang, and Z.~Han, ``A survey of emerging m2m systems: Context, task, and objective,'' \emph{IEEE Internet Things J.}, vol.~3, no.~6, pp. 1246--1258, Dec. 2016.

\bibitem{ref3}
L.~Chettri and R.~Bera, ``A comprehensive survey on internet of things (\uppercase{I}o\uppercase{T}) toward 5\uppercase{G} wireless systems,'' \emph{IEEE Internet Things J.}, vol.~7, no.~1, pp. 16--32, Jan. 2020.

\bibitem{ref4}
Z.~Chen and D.~Smith, ``Mmwave m2m networks: Improving delay performance of relaying,'' \emph{IEEE Wirel. Commun.}, vol.~20, no.~1, pp. 577--589, Jan. 2020.

\bibitem{akyildiz2008survey}
I.~F. Akyildiz, W.-Y. Lee, M.~C. Vuran, and S.~Mohanty, ``A survey on spectrum management in cognitive radio networks,'' \emph{IEEE Commun. Mag.}, vol.~46, no.~4, pp. 40--48, April 2008.

\bibitem{hamalainen2002uwb}
M.~Hamalainen, V.~Hovinen, R.~Tesi, J.~H. Iinatti, and M.~Latva-aho, ``On the uwb system coexistence with gsm900, umts/wcdma, and gps,'' \emph{IEEE Journal on Select. Areas Comm.}, vol.~20, no.~9, pp. 1712--1721, Dec. 2002.

\bibitem{naik2018coexistence}
G.~Naik, J.~Liu, and J.-M.~J. Park, ``Coexistence of wireless technologies in the 5 ghz bands: A survey of existing solutions and a roadmap for future research,'' \emph{IEEE Comm. Surveys Tutorials}, vol.~20, no.~3, pp. 1777--1798, March 2018.

\bibitem{sklivanitis2015all}
G.~Sklivanitis, E.~Demirors, A.~M. Gannon, S.~N. Batalama, D.~A. Pados, and T.~Melodia, ``All-spectrum cognitive channelization around narrowband and wideband primary stations,'' in \emph{Proc. GLOBECOM, San Diego, CA}, Dec. 2015, pp. 1--7.

\bibitem{tountas2018all}
K.~Tountas, G.~Sklivanitis, D.~A. Pados, and S.~N. Batalama, ``All-spectrum digital waveform design via bit flipping,'' in \emph{Proc. IEEE GLOBECOM, Abu Dhabi, United Arab Emirates}, Dec. 2018, pp. 1--6.

\bibitem{ding2013all}
L.~Ding, K.~Gao, T.~Melodia, S.~N. Batalama, D.~A. Pados, and J.~D. Matyjas, ``All-spectrum cognitive networking through joint distributed channelization and routing,'' \emph{IEEE Wirel. Commun.}, vol.~12, no.~11, pp. 5394--5405, Nov. 2013.

\bibitem{sklivanitis2017sparse}
G.~Sklivanitis, P.~P. Markopoulos, S.~N. Batalama, and D.~A. Pados, ``Sparse waveform design for all-spectrum channelization,'' in \emph{Proc. IEEE ICASSP, New Orleans, LA}, March 2017, pp. 3764--3768.

\bibitem{sklivanitis2018airborne}
G.~Sklivanitis, A.~Gannon, K.~Tountas, D.~A. Pados, S.~N. Batalama, S.~Reichhart, M.~Medley, N.~Thawdar, U.~Lee, J.~D. Matyjas \emph{et~al.}, ``Airborne cognitive networking: Design, development, and deployment,'' \emph{IEEE Access}, vol.~6, pp. 47\,217--47\,239, July 2018.

\bibitem{lee2008optimal}
W.-Y. Lee and I.~F. Akyildiz, ``Optimal spectrum sensing framework for cognitive radio networks,'' \emph{IEEE Wirel. Commun.}, vol.~7, no.~10, pp. 3845--3857, Oct. 2008.

\bibitem{kang2010optimal}
X.~Kang, H.~K. Garg, Y.-C. Liang, and R.~Zhang, ``Optimal power allocation for ofdm-based cognitive radio with new primary transmission protection criteria,'' \emph{IEEE Wirel. Commun.}, vol.~9, no.~6, pp. 2066--2075, June 2010.

\bibitem{rose2002wireless}
C.~Rose, S.~Ulukus, and R.~D. Yates, ``Wireless systems and interference avoidance,'' \emph{IEEE Wirel. Commun.}, vol.~1, no.~3, pp. 415--428, July 2002.

\bibitem{yates1995framework}
R.~D. Yates, ``A framework for uplink power control in cellular radio systems,'' \emph{IEEE Journal on Select. Areas Comm.}, vol.~13, no.~7, pp. 1341--1347, Sep. 1995.

\bibitem{tountas2019dynamic}
K.~Tountas, G.~Sklivanitis, and D.~A. Pados, ``Dynamic joint phy-mac waveform design for \uppercase{I}o\uppercase{T} connectivity,'' in \emph{Proc. IEEE ICASSP, Brighton, UK}, May 2019, pp. 8399--8403.

\bibitem{zhang2020prospective}
J.~Zhang, E.~Bj{\"o}rnson, M.~Matthaiou, D.~W.~K. Ng, H.~Yang, and D.~J. Love, ``Prospective multiple antenna technologies for beyond 5\uppercase{G},'' \emph{IEEE Journal on Select. Areas Comm.}, vol.~38, no.~8, pp. 1637--1660, June 2020.

\bibitem{abdulkadir2016space}
Y.~Abdulkadir, O.~Simpson, N.~Nwanekezie, and Y.~Sun, ``Space-time opportunistic interference alignment in cognitive radio networks,'' in \emph{Proc. IEEE WCNC, Doha, Qatar}, April 2016, pp. 1--6.

\bibitem{tountas2019directional}
K.~Tountas, G.~Sklivanitis, and D.~A. Pados, ``Directional space-time waveform design for interference-avoiding mimo configurations,'' in \emph{Proc. iWAT, Miami, FL}, March 2019, pp. 235--238.

\bibitem{liu2012joint}
Z.~Liu, D.~Sun, L.~Liu, and K.~Yi, ``Joint interference suppression for multi-user tdd mimo downlink with other-cell interference,'' \emph{IEEE Commun. Lett.}, vol.~16, no.~3, pp. 307--309, Jan. 2012.

\bibitem{zhang2008exploiting}
R.~Zhang and Y.-C. Liang, ``Exploiting multi-antennas for opportunistic spectrum sharing in cognitive radio networks,'' \emph{IEEE J. Sel. Top. Signal Process.}, vol.~2, no.~1, pp. 88--102, Feb. 2008.

\bibitem{osman2021novel}
R.~A. Osman, S.~N. Saleh, and Y.~N. Saleh, ``A novel interference avoidance based on a distributed deep learning model for 5\uppercase{G}-enabled \uppercase{I}o\uppercase{T},'' \emph{Sensors}, vol.~21, no.~19, p. 6555, Sep. 2021.

\end{thebibliography}

\end{document}